\documentclass[aps,prb,amsmath,amssymb
,showpacs
,floatfix
,twocolumn
]{revtex4}
\usepackage{graphicx}
\usepackage{dcolumn}
\usepackage{bm}
\usepackage{subfigure}
\begin{document}
\title{\bf Spin-Orbit Coupling in an f-electron Tight-Binding Model}

\author{M. D. Jones}
\affiliation{Department of Physics and Center for Computational Research,\\
University at Buffalo, The State University of New York, Buffalo, NY 14260}
\email{jonesm@ccr.buffalo.edu}

\author{R. C. Albers}

\affiliation{
Theoretical Division, Los Alamos National Laboratory, \\
Los Alamos, NM 87501}
\email{rca@lanl.gov}

\date{\today}

\begin{abstract}

We extend a tight-binding method to include the effects of spin-orbit coupling, 
and apply it to the study of the electronic properties of 
the actinide elements Th, U, and Pu.  These tight-binding parameters
are determined for the fcc crystal structure using the equivalent equilibrium
volumes.  
In terms of the single particle energies and the electronic density of
states, the overall quality of the tight-binding representation is excellent
and of the same quality as without spin-orbit coupling.
The values of the optimized tight-binding spin-orbit coupling parameters are 
comparable to those determined from purely atomic calculations.


\end{abstract}
%

\maketitle

\section{Introduction}
The accurate determination of inter-atomic forces is crucial for almost all aspects
of modeling the fundamental behavior of materials.
Whether one is interested in static equilibrium properties using Monte Carlo methods,
or time dependent phenomena using molecular dynamics, the essential feature remains
the origin, applicability, and transferability of the forces acting on the fundamental
unit being modeled (atoms or molecules in most cases).  
First principles methods
based on density functional theory have gained wide acceptance for their ease of
use, relatively accurate determination of fundamental properties, and high 
transferability.
These techniques, however, are limited in their application by current computing
technology to systems of a few hundred atoms or less (most commonly a few dozen atoms).
Potentials that are classically derived (i.e., pair potentials) lack directional
bonding (or at best add some bond angle information) and other quantum mechanical effects
but are computationally far more tractable for larger simulations.
Recent advances in tight-binding (TB) theory, which include directional bonding, but
treat only the most important valence electrons shells, therefore show a great deal of 
promise.

TB models have become a useful method for the computational modeling of
materials properties thanks to their ability to incorporate quantum mechanics
in a greatly simplified theoretical treatment, making large accurate simulations 
possible on modern digital computers\cite{goringe97,papa03}.  
Another advantage of these TB models is their ability to treat a general
class of problems that include directional bonding between valence electrons,
of particular importance for transition metal and $f$-electron materials.
Finally, TB models are widely used in many-body formalisms for the one-electron
part of the Hamiltonian.  It is therefore a useful representation of the band-structure
for a more sophisticated treatment of electronic correlation, and has so been used\cite{zhu07},
for example, in dynamical mean-field theory applications for Pu.

In this report we present recent developments towards a transferable tight-binding
total energy technique applicable to heavy metals.  With the addition of spin-orbit
coupling effects for angular momentum up to (and including) $f$-character, we 
demonstrate the applicability of this technique for the element Pu, of particular 
interest for its position near the half-filling point of the $5f$ subshell in the 
actinide sequence and the boundary between localized and delocalized 
$f$-electrons\cite{albers01}.

\section{TB Method}

The TB model used in this report is similar to that used in the handbook by
Papaconstantopoulos\cite{papacon86}.
We have extended the calculations to include $f$-electrons\cite{jones02} and spin-orbit
coupling\cite{lach-hab02}.  
As such, in this report we will elaborate only on those aspects of the technique 
that are unique to this work.
A very brief recapitulation of the underlying TB method and its approximations is
included to create the proper context for the addition of $f$-electrons and spin-orbit
coupling.

The Slater-Koster method\cite{slater54} consists of solving the secular
equation,
\begin{equation}
H\psi_{i,v}=\epsilon_{i,v} S\psi_{i,v},
\end{equation}
for the single-particle eigenvalues and orbitals, under the following restrictions:
terms involving more than two centers are ignored, terms where the
orbitals are on the same atomic site are taken as constants, and
the resulting reduced set of matrix elements are treated as variable
parameters.
The Hamiltonian, $H$, 
includes the labels for orbitals having generic quantum numbers $\alpha,
\beta$ localized on atoms $i,j$, where the effective potential is assumed
to be spherical, and can be represented as a sum over atomic centers,
\begin{equation}
H_{\alpha i,\beta j} = \left\langle \alpha,i \left| -\nabla^2
                        +\sum_k V^{\rm eff}_k \right| \beta,j \right\rangle,
\label{eq:hamsum}
\end{equation}
which we further decompose into ``on-site'' and ``inter-site'' terms,
\begin{equation}
H_{\alpha i,\beta j} = e_\alpha \delta_{\alpha\beta} \delta_{ij}
        + E_{\alpha i,\beta j\ne i},
\end{equation}
where the on-site terms, $e_\alpha$, represent terms in which two
orbitals share the same atomic site, and
\begin{widetext}
\begin{equation}
E_{\alpha i,\beta j\ne i} = \sum_n e^{i \textbf{k}\cdot\left(\textbf{R}_n+
\textbf{b}_j-\textbf{b}_i\right)} \int d\textbf{r} \psi_\alpha\left(
        \textbf{r}-\textbf{R}_n-\textbf{b}_i\right) H
        \psi_\beta\left( \textbf{r}-\textbf{b}_j\right),
\label{eq:enint}
\end{equation}
\end{widetext}
are the remaining energy integrals involving orbitals located
on different atomic sites, and we have used translational
invariance to reduce the number of sums over bravais lattice
points $\{\textbf{R}_n\}$, and the $\textbf{b}_i$ denote atomic
basis vectors within the repeated lattice cells.
Note that terms which have both orbitals located on the same site,
but the effective potential ($V^{\textrm{eff}}$) on other sites have been ignored.
These contributions
are typically taken to be ``environmental'' corrections to the
on-site terms, and are not accounted for in the usual Slater-Koster formalism.
For the inter-site terms, the two center approximation also consists of ignoring 
these additional terms in which the effective potential, $V^{\textrm{eff}}$, 
does not lie on one of the atomic sites.
Once this approximation has been made,
the inter-atomic ($i\ne j$) matrix elements reduce to a simple
sum over angular functions, $G_{ll'm}(\Omega_{i,j})$, and functions which 
depend only upon the magnitude of the distances between atoms,
\begin{equation}
H_{\alpha i,\beta j} = \sum h_{ll'm}(r_{ij}) G_{ll'm}(\Omega_{i,j}),
\label{eq:mat2}
\end{equation}
where we have now adopted the usual convention of using
the familiar $l,m$ angular momentum quantum numbers,
and the axis connecting the atoms is the quantization axis.
An equivalent expression for $s_{ll'm}$ terms exists when non-orthogonal orbitals
are used.
The basis set used for the $\alpha$ and $\beta$ quantum states are
the cubic harmonics\cite{vonderlage47} 
whose functional forms are given in Table \ref{tab:TBbasis} (with appropriate normalization 
factors)
\squeezetable
\begin{table*}[ht]
\begin{ruledtabular}
\caption{TB basis functions used for an $sp^3d^5f^7$ calculation.  Note that
$f_l(r)=1/r^l$.}\label{tab:TBbasis}
\begin{tabular}{llllllll}
\multicolumn{2}{c}{l=0}&\multicolumn{2}{c}{l=1}&\multicolumn{2}{c}{l=2}&
\multicolumn{2}{c}{l=3}\\ 
\cline{1-2}\cline{3-4}\cline{5-6}\cline{7-8}
$| s\pm \rangle=$ & $\sqrt{1/4\pi} |\pm\rangle$ & $| p_1\pm \rangle=$ 
                 & $\sqrt{3/4\pi} f_1(r) x |\pm\rangle $ 
                 & $| d_1\pm \rangle=$ & $\sqrt{5/16\pi} f_2(r) xy |\pm\rangle$ 
                 & $| f_1\pm \rangle=$ & $2\sqrt{105/16\pi} f_3(r) xyz |\pm\rangle$ \\
&&$| p_2\pm \rangle=$ & $\sqrt{3/4\pi} f_1(r) y |\pm\rangle $ 
		&$| d_2\pm \rangle=$ & $2\sqrt{15/16\pi} f_2(r) yz |\pm\rangle$ 
		&$| f_2\pm \rangle=$ & $\sqrt{7/16\pi} f_3(r) x(5x^2-3r^2) |\pm\rangle$ \\
&&$| p_3\pm \rangle=$ & $\sqrt{3/4\pi} f_1(r) z |\pm\rangle $
		&$| d_3\pm \rangle=$ & $2\sqrt{15/16\pi} f_2(r) zx |\pm\rangle$ 
		&$| f_3\pm \rangle=$ & $\sqrt{7/16\pi} f_3(r) y(5y^2-3r^2) |\pm\rangle$ \\
&&&&$| d_4\pm \rangle=$ & $\sqrt{15/16\pi} f_2(r) (x^2-y^2) |\pm\rangle$ 
		&$| f_4\pm \rangle=$ & $\sqrt{7/16\pi} f_3(r) z(5z^2-3r^2) |\pm\rangle$ \\
&&&&$| d_5\pm \rangle=$ & $\sqrt{5/16\pi} f_2(r) (3z^2-r^2) |\pm\rangle$ 
		&$| f_5\pm \rangle=$ & $\sqrt{105/16\pi} f_3(r) x(y^2-z^2) |\pm\rangle$ \\
&&&&&&$| f_6\pm \rangle=$ & $\sqrt{105/16\pi} f_3(r) y(z^2-x^2) |\pm\rangle$ \\
&&&&&&$| f_7\pm \rangle=$ & $\sqrt{105/16\pi} f_3(r) z(x^2-y^2) |\pm\rangle$
\end{tabular}
\end{ruledtabular}
\end{table*}
where $|\pm\rangle$ denotes the spin-state, which
we will need for spin-orbit coupling.

The Slater-Koster tables for the $sp^3d^5$ matrix elements can be
found in standard references\cite{harrison80}, and we have used
the tabulated results of Takegahara {\it et al.}\cite{takegahara80} for the
additional matrix elements involving $f$-electrons.
Typical TB applications are then reduced to using TB as an interpolation
scheme; the matrix elements ($h_{ll'm}$, $s_{ll'm}$ and 
$e_\alpha$) are
determined by fitting to {\it ab-initio} calculated quantities such as the
total energy and band energies.

In this study we restrict ourselves to the determination of optimal TB
parameters at the neighbor distances in the face-centered cubic
crystal structure (often used as a surrogate for the more complex ground
state crystal structure of the actinides) near the equilibrium volume.
Such tabulations have been extensively used\cite{papacon86} in the study
of materials with lower atomic number.  To the best of our knowledge this
is the first time that such parameters have been presented for light actinide
elements that include the $f$-electron orbitals (although similar parameters 
have been
determined for the elements Ac and Th in an $sp^3d^5$ basis\cite{papacon86}).
The TB parameter values so derived are available (on request) from the
authors.



\subsection{Spin-orbit coupling}

The primary impact of spin-orbit coupling is to non-trivially couple
electrons of different spin states, thus doubling the size of the TB
Hamiltonian.
The spin-orbit contribution to the Hamiltonian is given by
\begin{equation}
\label{eq:Hso} 
H^{so} = \xi(r)\textbf{L}\cdot\textbf{S},
\end{equation}
where 
$\xi(r)=(\alpha^2/(2r))(\partial V/\partial r)$, 
$V$ is the total
(crystal) potential.  We neglect contributions from more than one
center.
A new Hamiltonian matrix can then be defined in terms of the spinless
one,
\begin{equation}
{\cal H} = H+H^{so} = \left( 
\begin{array}{cc} H + \frac{1}{2} \xi L_z & \frac{1}{2} \xi L_- \\
		    \frac{1}{2} \xi L_+ & H-\frac{1}{2} \xi L_z
\end{array}
\right)
\end{equation}
where 
\begin{equation}
\xi_{nl} =  \hbar \int_0^\infty \xi(r) \left[ R^0_{nl}(r)\right]^2 r^2dr,
\end{equation}
is the spin-orbit coupling parameter between orbitals
of orbital angular momentum $l$ and primary quantum number $n$ located on 
the same atom, $L_\pm$
are the usual raising and lowering operators, and $L_z$ the azimuthal
angular momentum operator,
\begin{equation}
\begin{array}{ll}
L_\pm Y_{lm} (\theta,\phi) = \hbar\sqrt{l(l+1)-m(m\pm 1)} Y_{lm\pm 1} \nonumber \\
L_z Y_{lm} (\theta,\phi) = \hbar m Y_{lm}.
\end{array}
\end{equation}
The functions $R^0_{nl}(r)$ are the non-relativistic radial wave functions.
The spin-orbit contributions to the Hamiltonian matrix can then be
expressed in term of the TB basis functions listed in Table \ref{tab:TBbasis}.
Rather than list contributions for the 32x32 matrix, here we list the 
matrices in the sub-blocks corresponding to each orbital angular
momentum.  The $p$ and $d$ contributions have been previously discussed
in relation to the tight-binding formalism\cite{friedel64,chadi77}; to the
best of our knowledge no $f$ contribution has yet appeared in the literature.
For completeness we detail the spin-orbit contribution for all values of
the angular momentum up to $l=3$.
\begin{widetext}
\begin{equation}
\label{eq:hso_p}
{H}^{so}_p = \frac{\xi_{np}}{2}\left(
\begin{array}{rrrrrr}
0 & -i & 0  & 0 & 0 & 1 \\
i & 0 & 0 & 0 & 0 & -i \\
0 & 0 & 0 & -1 & i & 0 \\
0 & 0 & -1 & 0 & i & 0 \\
0 & 0 & -i & -i & 0 &0 \\
1 & i & 0 & 0 & 0 & 0 
\end{array}\right),
\end{equation}

\begin{equation}
\label{eq:hso_d}
{H}^{so}_d = \frac{\xi_{nd}}{2}\left(
\begin{array}{rrrrrrrrrr}
0  & 0  & 0  & 2i & 0   & 0  & 1         &-i & 0 & 0  \\
0  & 0  & i  & 0  & 0   &-1  & 0         & 0 &-i &-i\sqrt{3} \\
0  & -i & 0  & 0  & 0   & i  & 0         & 0 &-1 & \sqrt{3} \\
-2i& 0  & 0  & 0  & 0   & 0  &  i        &  1& 0 & 0 \\
0  & 0  & 0  & 0  & 0   & 0  &  i\sqrt{3}&-\sqrt{3} & 0 & 0 \\
0  &-1  &-i  & 0   & 0   & 0  & 0         & 0        & -2i & 0 \\
 1 & 0  & 0 & -i   &-i\sqrt{3} & 0 & 0 & -i& 0 & 0 \\
 i & 0  & 0 &  1   &-\sqrt{3}  & 0 & i& 0 & 0 & 0 \\
0 &  i  & -1& 0    &  0        & 2i& 0 & 0 & 0 & 0 \\
0 & i\sqrt{3} & \sqrt{3} & 0 & 0  & 0 & 0 & 0 & 0 & 0 
\end{array}
\right),
\end{equation}

\begin{equation}
\label{eq:hso_f}
{H}^{so}_f = \frac{\xi_{nf}}{4}\left(
\begin{array}{rrrrrrrrrrrrrr}
0  & 0     & 0    & 0 & 0  & 0  & 2i   & 0 & 0 & 0   & 0  & 2i & 2 & 0   \\
0  & 0     & \frac{3i}{2} & 0 & 0  & it & 0    & 0 & 0 & 0   &-\frac{3}{2}& 0  & 0 & t   \\
0  & -\frac{3i}{2} & 0    & 0 & it & 0  & 0    & 0 & 0 & 0   &\frac{3i}{2}& 0  & 0 & it  \\
0  & 0     & 0    & 0 & 0  & 0  & 0    & 0 &\frac{3}{2}&-\frac{3i}{2}& 0  & t  &it & 0   \\
0  & 0     & -it  & 0 & 0  &-\frac{i}{2}& 0    &-2i& 0 & 0   & -t & 0  & 0 &\frac{1}{2} \\
0  & -it   & 0    & 0 &\frac{i}{2}& 0  & 0    &-2 & 0 & 0   & -it& 0  & 0 & -\frac{i}{2} \\
-2i& 0     & 0    & 0 & 0  & 0  & 0    & 0 &-t &-it  & 0  &-\frac{1}{2}&\frac{i}{2}& 0    \\
0  & 0 & 0  & 0   &2i& -2 & 0          & 0 & 0  & 0   & 0 & 0  & 0 & -2i  \\
0  & 0 & 0  &\frac{3}{2} & 0 & 0 & -t         & 0 & 0  &-\frac{3i}{2}& 0 & 0  &-it& 0 \\
0  & 0 & 0  &\frac{3i}{2}& 0 & 0 & it        & 0 &\frac{3i}{2}& 0   & 0 &-it & 0 & 0  \\
0  &-\frac{3}{2}&-\frac{3i}{2}& 0   & -t &it& 0          & 0 & 0  & 0   & 0 & 0  & 0 & 0  \\
-2i& 0 & 0  & t   & 0 & 0 & -\frac{1}{2}       & 0 & 0  & it  & 0 & 0  &\frac{i}{2}& 0  \\
2  & 0 & 0  &-it  & 0 & 0 & -\frac{i}{2}       & 0 & it & 0   & 0 &-\frac{i}{2}& 0 & 0 \\
0  & t & -it& 0   &\frac{1}{2}&\frac{i}{2}& 0          & 2i& 0  & 0   & 0 & 0  & 0 & 0
\end{array}
\right),
\end{equation}
\end{widetext}
where $t=\sqrt{15}/2$.

\subsection{Fitting the Parameters}

\begin{figure}[th!]
\centering\includegraphics[scale=0.3,angle=-90]{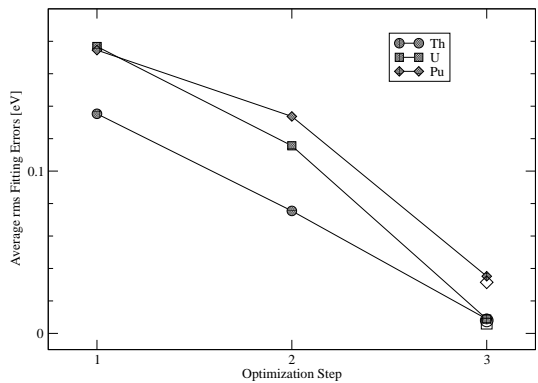}
\caption{TB fit quality in terms of the cumulative root mean square (rms) errors 
at various steps of the optimization procedure. Step 1 relaxes the spin-orbit
parametes ($\xi_{nl}$), 2 relaxes the remaining on-site parameters, and 3 is a full
relaxation of all parameters.  Open symbols at Step 3 indicate the original
scalar-relativistic fit quality.  Note that the cumulative rms error is over all of the
fitted bands (20 bands for Th, U, and Pu).
Although the spin-orbit coupling is an atomic quantity, the improvement of
our results in step 3 (which relaxes inter-site parameters) indicates some
environmental effects should also be taken into account.
}\label{fig:rms}
\end{figure}

The values of the TB parameters were determined using standard non-linear least 
squares optimization routines by matching energy band values derived
from highly accurate first principles density functional theory (DFT)
calculations\cite{wien2k}.  The technique is described in detail in 
a previous work\cite{jones02}, where the DFT calculations in this case used
a generalized gradient approximation DFT functional\cite{pbe96}, and the
improved tetrahedron scheme\cite{blochl94} for Brillouin zone integrations.
In this study we use as a starting point high 
quality fits to the scalar-relativistic energy bands and approximate atomic values of
the spin-orbit parameters.  The first step  
is to then use this fit for fitting the relativistic energy bands including spin-orbit
coupling.  Successive optimization steps then relax only the spin-orbit coupling paramaters
(step 1), the remaining on-site parameters (step 2), and finally the 
inter-site terms (step 3).  The fit quality through these steps is shown in Figure
\ref{fig:rms}.  Note that the quality of the final fit is comparable to the original
fit quality (open symbols at step 3) when only scalar-relativistic effects were
taken into account.

\section{Application to the Light Actinides, Th, U, and Pu}
\subsection{Energy bands including spin-orbit coupling}

\begin{figure}[!ht]
\begin{center}
 \subfigure[Th]{\includegraphics[scale=0.25]{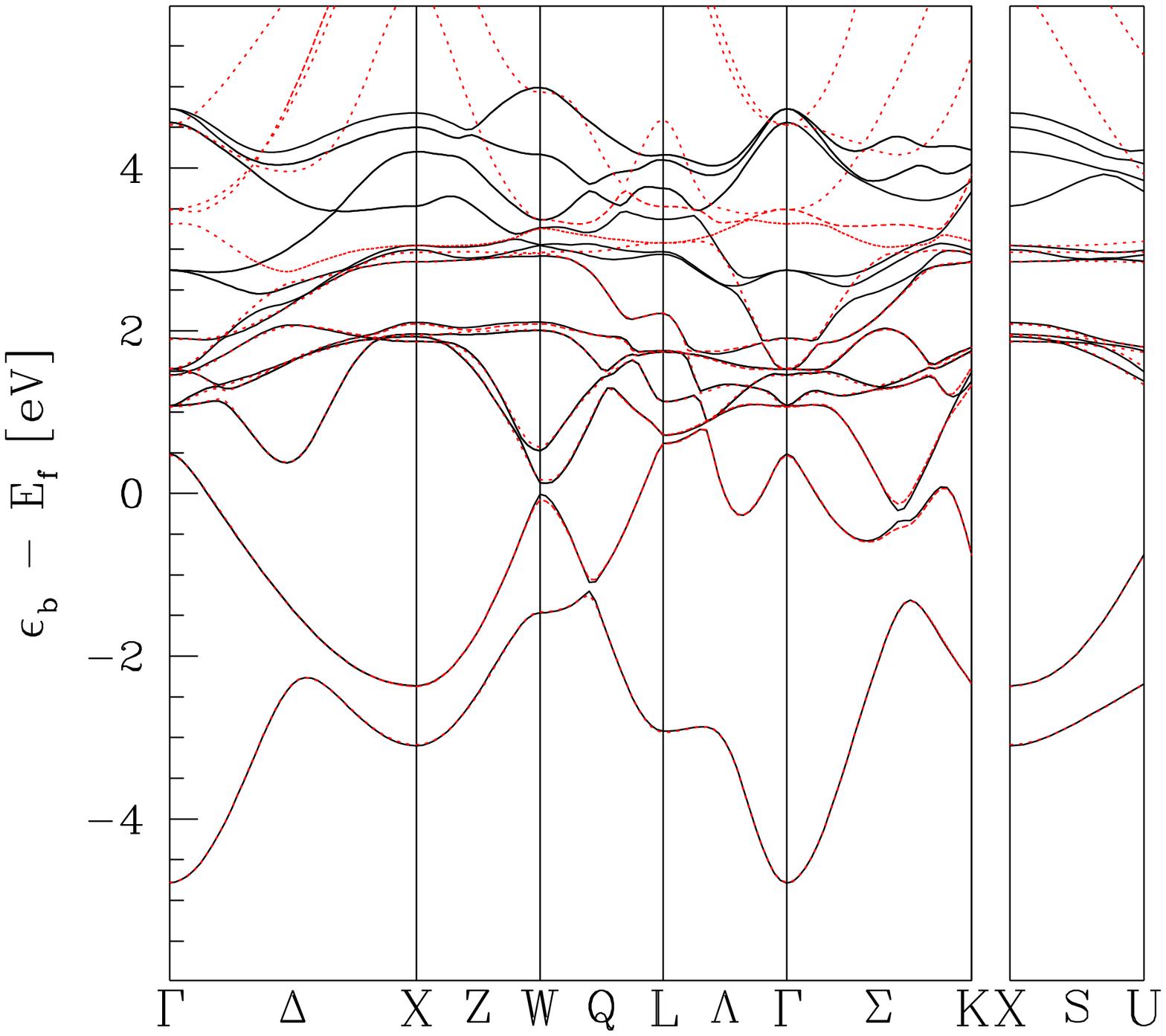}}
 \subfigure[U]{\includegraphics[scale=0.25]{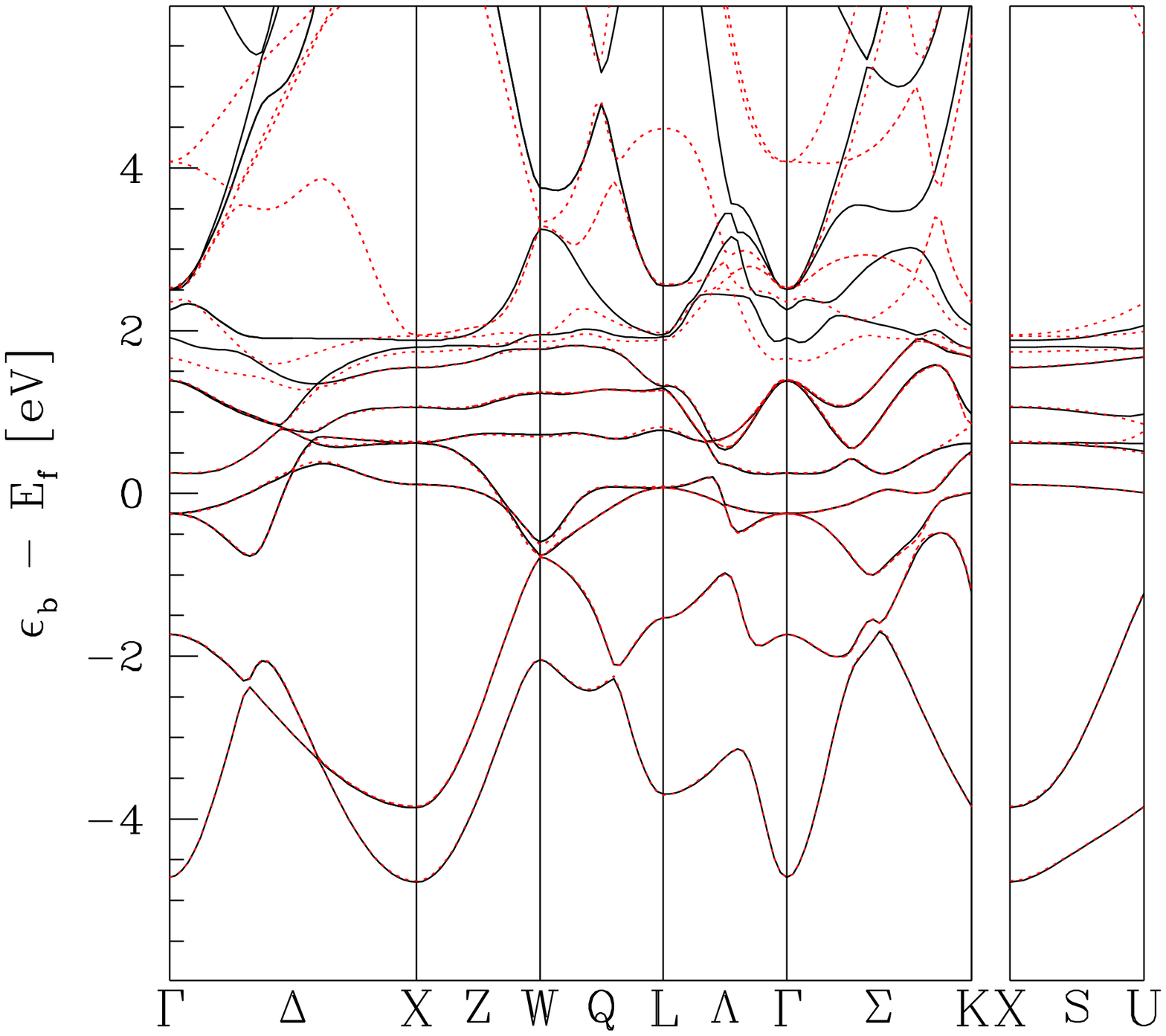}} 
 \subfigure[Pu]{\includegraphics[scale=0.25]{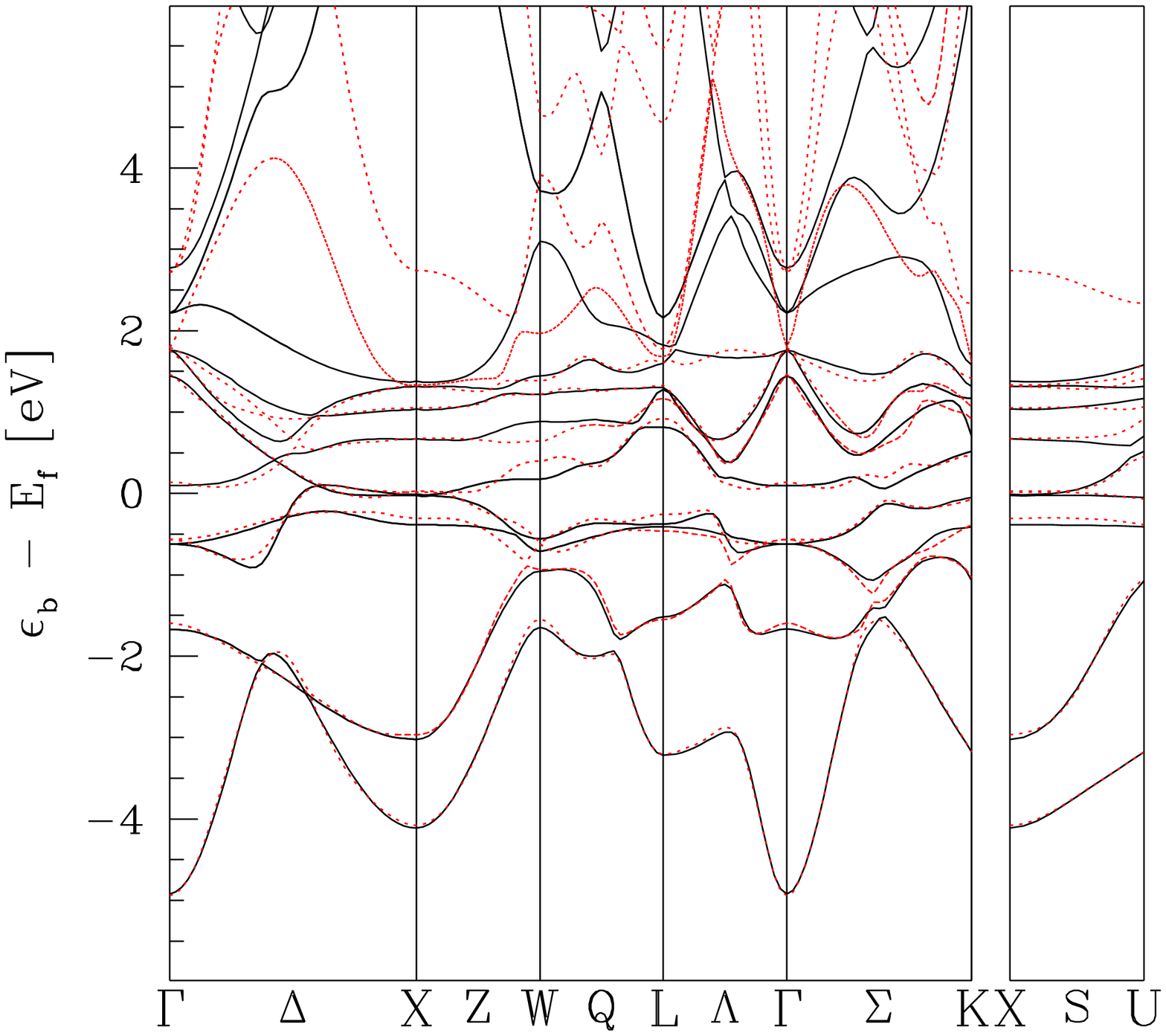}}
\end{center}
\caption{TB energy bands for Th ($a=9.61$), U ($a=8.22$), and Pu ($a=8.14)$, shown 
in comparison with FLAPW valence energy bands (dotted lines). Note the excellent 
agreement.  The abscissa for each calculation has been shifted
such that the Fermi energy is at zero.  Higher valence states (above
the first 20) are not fit, hence the poorer fit quality well above
the Fermi level.}\label{fig:pu-bands}
\end{figure}

The first comparison between the TB fit and FLAPW calculations are the energy bands
shown in Figure \ref{fig:pu-bands}.  Note the excellent agreement between the two sets
of calculations (the cumulative root mean square error in the TB fits to the first 20 
energy bands in the irreducible Brillouin zone is 0.013, 0.013, and 0.072 Ry, 
respectively).  
\begin{figure}[!ht]
\begin{center}
\includegraphics[scale=0.25]{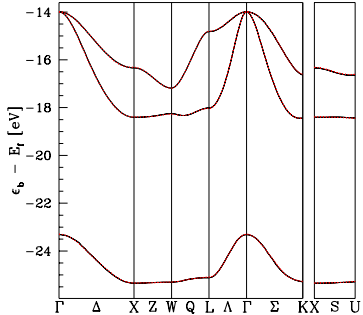}
\end{center}
\caption{TB energy bands (dashed lines) for Pu semi-core 6p states, compared with 
FLAPW values (solid lines).}\label{fig:pu-bands6p}
\end{figure}
Also note that we have included 
the ``semi-core'' $6p$ states in the fit to better fix the available $p$ states in the 
TB basis.  To expand the energy scale comparing the valence bands, the fit quality for
the semi-core 6p states is shown separately in Figure \ref{fig:pu-bands6p} for Pu
(all three elements have similar excellent fit quality for the more localized 6p 
states).
Note that higher energy bands (well above the Fermi level) are not fit, hence
the larger discrepancies for those levels.

\subsection{Density of states including spin-orbit coupling} 
\begin{figure}[!ht]
\begin{center}
 \subfigure[Th]{\includegraphics[scale=0.25,angle=-90]{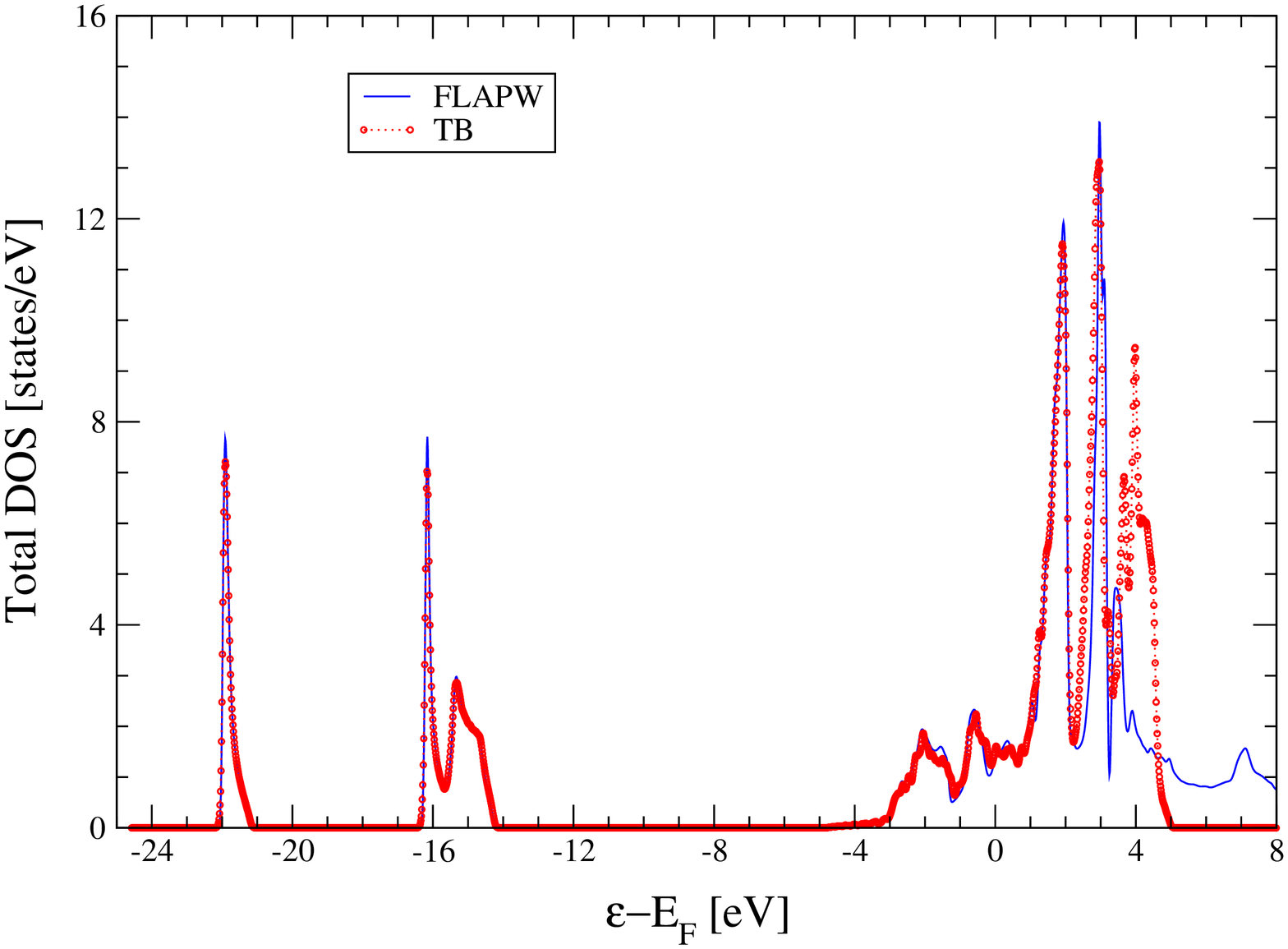}} \quad
 \subfigure[U]{ \includegraphics[scale=0.25,angle=-90]{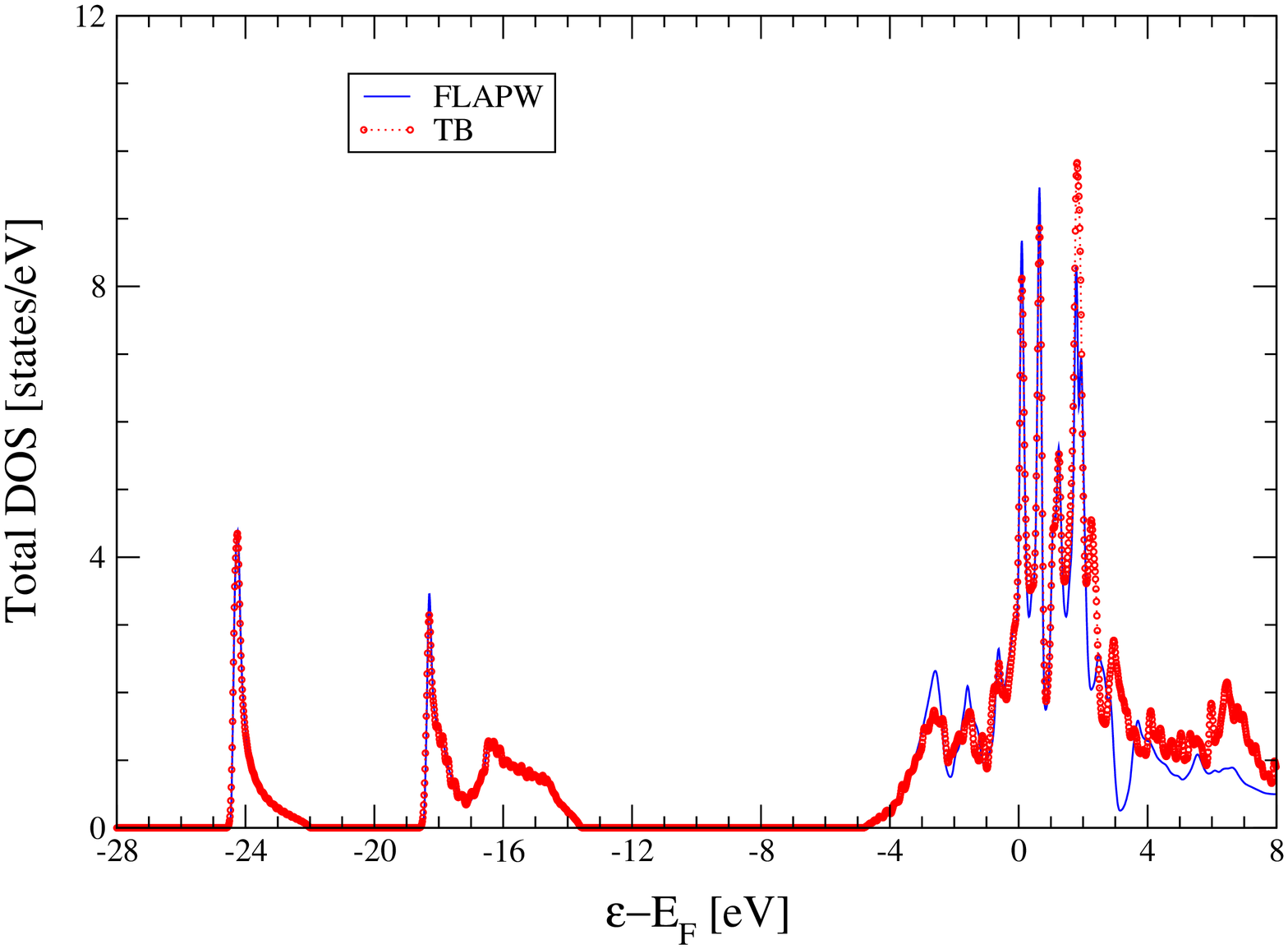}}  \quad 
 \subfigure[Pu]{\includegraphics[scale=0.25,angle=-90]{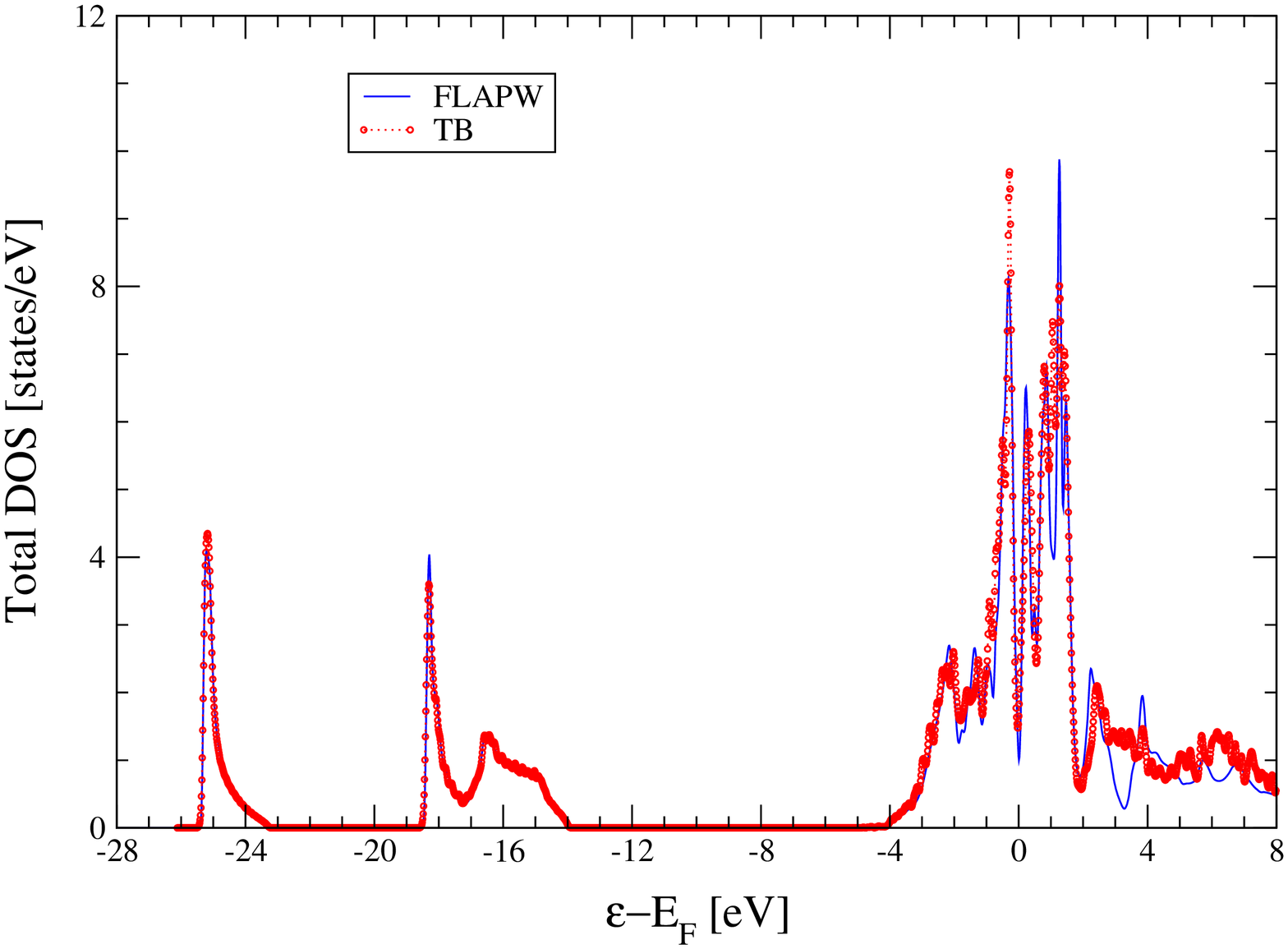}}
\end{center}
\caption{TB (dotted lines) and FLAPW (solid lines) total DOS, including 
spin-orbit coupling.  
Note that
the TB calculation is in quite good agreement with the FLAPW results, despite using a
different BZ integration method.The abscissa for each calculation has been shifted
such that the Fermi energy is at zero.}\label{fig:pu-so-tdos}
\end{figure}

We also compare the total density of states (DOS) between
TB and FLAPW methods in Figure \ref{fig:pu-so-tdos}.  

The TB
method shown in the figure used a simple Fermi-Dirac temperature smearing method
(with $k_BT=500$)
for integrating over the irreducible wedge of the Brillouin zone, while the FLAPW
calculations used the improved tetrahedron\cite{blochl94} method with Gaussian smearing.
From the comparison between the TB and FLAPW methods shown in the above figure, we 
note that the agreement is excellent, with all major features in the DOS reproduced
by the TB calculations.  
There is a slight reduction in the height of some of the larger peaks in the DOS
for the TB technique, most likely due to the inability of the temperature smearing
technique to represent the finer grained features as well as the improved tetrahedron
method does.


\subsection{Spin-orbit coupling terms}  

\begin{table*}[!htb]
\caption{\label{tab:split}
Values of spin-orbit coupling strength, $\xi_{nl}$, and spin-orbit splittings, 
$\Delta_{nl}= (2l+1)\xi_{nl}/2$, for the various valence
electron shells predicted by the TB fit compared with purely atomic
values using relativistic density functional theory (DFT)\cite{kotochigova96},
a Dirac-Slater atomic code (DIRAC)\cite{adf},
and relativistic Hartree-Fock-Slater (HFS)\cite{herman63} atomic calculations.
Dashed entries are used for orbitals not populated in the
atomic calculations.  Values are in eV.}
\begin{ruledtabular}
\begin{tabular}{lllllll}
\hline
Method & $\xi_{6p}$ & $\Delta_{6p}$ & $\xi_{5d}$ & $\Delta_{5d}$ & $\xi_{5f}$ & $\Delta_{5f}$ \\
&\multicolumn{6}{c}{Th}\\
DIRAC  & 5.29 & 7.94 & 0.20 & 0.51 & 0.19 & 0.66 \\
DFT    & 5.24 & 7.86 & 0.21 & 0.52 & --   & --\\
HFS    & 4.09 & 6.14 & 0.30 & 0.75 & --   & -- \\
TB     & 4.19 & 6.29 & 0.20 & 0.51 & 0.18 & 0.62 \\
&\multicolumn{6}{c}{U}\\
DIRAC  & 5.96 & 8.94 & 0.19 & 0.47 & 0.24 & 0.83 \\
DFT    & 5.90 & 8.85 & 0.20 & 0.50 & 0.24 & 0.84 \\
HFS    & 4.38 & 6.57 & 0.30 & 0.75 & 0.35 & 1.24 \\
TB     & 4.64 & 6.96 & 0.23 & 0.58 & 0.42 & 1.48 \\
&\multicolumn{6}{c}{Pu}\\
DIRAC  & 6.92 & 10.38 & 0.20 & 0.51 & 0.31 & 1.10 \\
DFT    &  --  & --    & --   & --   & --   & --  \\
HFS    & 4.60 & 6.90  & --   & --   & 0.41 & 1.43 \\
TB     & 5.23 & 7.84  & 0.59 & 1.46 & 0.54 & 1.90 \\
\hline
\end{tabular}
\end{ruledtabular}
\end{table*}
It is interesting to compare the spin-orbit
coupling parameters, $\xi_{nl}$, predicted by TB theory for the various valence shells
relative to the values predicted by accurate Hartree-Fock-Slater calculations of
isolated atoms\cite{herman63}.  This comparison is shown in Table \ref{tab:split}.


Note the overall agreement between the TB fitted parameters and the atomic values.
The overall shift of a few tenths of an eV for the TB values is interesting,
and this trend could be representative of crystal field effects (this speculation
could be checked by performing equivalent fits at different densities).
Equivalently, one can compare the spin-orbit splitting of the electronic energy
levels with the purely atomic case.  This comparison is also shown in 
Table \ref{tab:split}.

\section{Conclusions}
We have included $f$-electron and spin-orbit effects in a standard tight-binding 
method for solids in order to advance simpler simulation methods that are capable 
of the accuracy of more expensive, full-potential density-functional techniques.
We have applied this TB technique to elemental fcc Th, U, and Pu, and have
achieved excellent agreement
with the electronic properties predicted using a highly accurate FLAPW method.
The fitted spin-orbit coupling parameters match very well the values
independently predicted by atomic electronic structure calculations.
This methodology bodes well for further TB investigations, especially
for the study of defects, phonons, and dynamical properties.  In future work
we intend to develop a more transferable model based on a TB total energy 
formalism\cite{jones02}, which should allow the straightforward calculation
of detailed materials properties.

\begin{acknowledgments}
This work was carried out under the auspices of the National Nuclear Security Administration
of the U.S. Department of Energy at Los Alamos National Laboratory under Contract No. DE-AC52-06NA25396.
Calculations were performed at
the Los Alamos National Laboratory and the Center for Computational
Research at SUNY--Buffalo.
FLAPW calculations were performed using the Wien2k package\cite{wien2k}.
We thank Jian-Xin Zhu for providing helpful remarks.
\end{acknowledgments}

%
\bibliography{TBSO_2col}


\end{document}